\documentclass[aps,prd,reprint,groupedaddress,showpacs]{revtex4-1}
\usepackage{amssymb}
\usepackage{graphicx}
\usepackage{dcolumn}

\begin{document}

\title{Resonant phenomena in finite motions of test particles in oscillating dark matter configurations}
\author{Vladimir A. Koutvitsky}
\author{Eugene M. Maslov}
\email{eugen.masloff2014@yandex.ru}
\affiliation{Pushkov Institute of Terrestrial Magnetism, Ionosphere and Radio Wave Propagation (IZMIRAN) 
of the Russian Academy of Sciences,\\ Moscow, Troitsk, Kaluzhskoe Hwy 4, Russian Federation, 108840}
\date{\today}

\begin{abstract}
Nonlinear differential equations are derived that describe the time
evolution of the test particle coordinates during finite motions in the
gravitational field of oscillating dark matter. It is shown that in the weak
field approximation, the radial oscillations of a test particle and
oscillations in orbital motion are described by the Hill equation and the
nonhomogeneous Hill equation, respectively. In the case of scalar dark matter
with a logarithmic self-interactions, these equations are integrated
numerically, and the solutions are compared with the corresponding solutions
of the original nonlinear system to identify possible resonance effects.
\end{abstract}

\maketitle

\section{Introduction}

The nature of dark matter (DM) remains one of the main mysteries of modern
cosmology. Currently, the most realistic hypothesis seems to be that DM
consists of ultralight bosons, e.g., axions \cite{Marsh}, with a mass in the 
$10^{-23}-10^{-21}$ eV range. New arguments in favor of this hypothesis are
provided by very recent astronomical observations \cite{Amruth}. Coherent
states of such ultralight particles are described by a classical scalar
field, whose wave properties solve serious problems of the standard $\Lambda 
$CDM model on galactic and subgalactic scales \cite{Turner, Seidel1,
Seidel2, Kolb, Lee, Peebles, Hu, Matos, Magana, Hui}. 
In the early Universe, the primordial fluctuations of this field are stretched 
by inflation to  form an almost homogeneous scalar background.
This background oscillates near the minimum of the effective potential with a
fundamental frequency $\omega $ depending on both the mass of the scalar
field and the shape of the potential. However, these oscillations are
unstable \cite{Khlopov}. If the bosons of DM are not self-interacting, the
corresponding free scalar field behaves on average as a dust-like matter 
\cite{Turner}. In this case, due to Jeans instability, the scalar background
breaks into diffuse lumps, which begin to collapse under the influence of
self-gravity. On a scale comparable to the Jeans length, the collapse of a
single lump stops due to the action of the so-called quantum pressure which
appears from the growing gradient of the energy density of the scalar field.
From the quantum mechanical point of view, this pressure is a manifestation
of the Heisenberg uncertainty principle. The result of the established
equilibrium is a "fuzzy" 
DM halo, in which the scalar field oscillates with the fundamental frequency 
$\omega\sim m$ \cite{Hu}. In addition, as numerical simulations show, the central
region of the formed halo, the core, undergoes quasi-normal low-frequency
oscillations, which are superimposed on the fundamental oscillations leading
to their modulation \cite{Veltmaat, Li}.

If DM particles are self-interacting, another scenario for the formation of
the localized DM configurations is realized. Self-interaction is described by
additional higher-order terms in the effective potential. These terms can be
either regular or singular. In the first case, they have a significant
effect on the dynamics of the scalar field only at large oscillation
amplitudes. In the second case, this is not necessary. An example of a
singular potential will be considered in Section III of our paper. In both
cases, the scalar field oscillations turn out to be 
unstable due to parametric resonance between the oscillating background and
perturbations. As a result, the homogeneous scalar background breaks up into
an ensemble of oscillating lumps, oscillons (pulsons). This mechanism works
on both cosmological and astrophysical scales \cite{Kofman, Koutvitsky0,
Amin, Zhang, Lozanov, Koutvitsky1, Fukunaga, Arvanitaki} (see \cite{Olle}
for a review). Under the action of gravity, after the completion of some
relaxation processes, oscillons turn into long-lived self-gravitating
oscillating objects with the size $\sim m^{-1}$, oscillatons (gravipulsons) 
\cite{Urena, Fodor, Koutvitsky2}, separated from the Hubble flow. This means
that each individual oscillaton is the result of an established dynamic
equilibrium between self-interaction, self-gravity and quantum pressure and
therefore must be described by the self-consistent system of
Einstein-Klein-Gordon equations (see \cite{Visinelli}\ for a recent review).
Note that, within this system of equations, oscillatons can also arise from rather
arbitrary localized initial conditions due to the gravitational cooling
process \cite{Seidel1, Seidel2, Guzman}.\ 

In any case, all the DM oscillations mentioned above lead to the
corresponding oscillations of the gravitational field, which can be detected
by their effect on the motion of photons and test particles. For example, as
shown in \cite{Khmelnitsky}, the gravitational time delay for a photon
passing through an oscillating halo should cause small periodic fluctuations
in the observed timing array of a pulsar located inside the halo, which can
be detected in future pulsar timing experiments. In paper \cite{Aoki}, it
was proposed to use the laser interferometers for detecting the
gravitational waves caused by the motion of the Earth through oscillating
DM. The secular variations of the orbital period in the system of binary
pulsars were discussed in \cite{Blas, Rozner} as a probe for ultralight
oscillating DM. Also, in this context, in Refs. \cite{Becerril, Boskovic}
the motion of test bodies in spherically symmetric time-periodic spacetimes
induced by a non-self-interacting DM was numerically investigated. In paper 
\cite{Boskovic} it was shown, in particular, that the orbital resonances may
occur in the motion of stars in oscillating halos. In addition, in 
\cite{Boskovic} it was demonstrated that spectroscopic emission lines from stars
in such halos exhibit characteristic, periodic modulations due to variations
in the gravitational frequency shift. These modulations were found
analytically in Ref. \cite{Koutvitsky3}.

Recently, in the above context, we studied the infinite motions of photons
and test particles in the oscillating DM \cite{Koutvitsky4, Koutvitsky4a}.
Namely, we considered deflection of their trajectories when passing through
an oscillating spherically symmetric lump of DM. Using the geodesic method and
the perturbative approach, we have found and calculated analytically
periodic variations in the deflection angle of both light rays and
trajectories of the test particles. In the present paper, we apply the same
approach to study the finite motions of the test particles in oscillating
DM, focusing on radial and orbital trajectories.

Our paper is organized as follows. In Sec. II, on the basis of geodesic
equations, we derive exact equations that describe the time dependence of
the radial and angular coordinates of a test particle making a finite motion
inside an oscillating self-gravitating lump of DM. We show that in the weak
field approximation these equations reduce to the homogeneous and
nonhomogeneous Hill equations for radial and perturbed circular motions,
respectively. In Sec. III we use these equations to describe the radial and
perturbed circular motions in the case when the DM lump is formed by a real
scalar field with a logarithmic self-interaction. Applying numerical
integration, we compare solutions to Hill equations with solutions to exact
equations to explore possible resonance effects. Discussion and concluding
remarks can be found in Sec. IV.

\section{Finite trajectories of test particles in time-periodic spherically
symmetric spacetimes}

The motion of a particle in a gravitational field obeys the geodesic equation%
\begin{equation}\label{eq1a}
\frac{d^{2}x^{\mu }}{ds^{2}}+\Gamma _{\alpha \beta }^{\mu }\frac{dx^{\alpha }%
}{ds}\frac{dx^{\beta }}{ds}=0,  
\end{equation}%
where $ds$ is the proper time interval. Consider a spherically symmetric
oscillating localized lump of DM. The gravitational field inside the lump is
described by the metric%
\begin{equation}\label{eq1b}
ds^{2}=B(t,r)\,dt^{2}-A(t,r)\,dr^{2}-r^{2}(d\vartheta ^{2}+\sin
^{2}\vartheta \,d\varphi ^{2}),  
\end{equation}%
where $A(t,r)$ and $B(t,r)$ are time-periodic and tend to unity as $%
r\rightarrow \infty $. For the trajectories lying in the plane $\vartheta
=\pi /2$, the geodesic equation reduces to the system%
\begin{equation}\label{eq2}
\frac{d}{ds}\ln \left( B\frac{dt}{ds}\right) =\frac{\dot{B}}{2B}\frac{dt}{ds}%
-\frac{\dot{A}}{2B}\left( \frac{dr}{ds}\right) ^{2}\left( \frac{dt}{ds}%
\right) ^{-1},  
\end{equation}%
\begin{equation}
\frac{d^{2}r}{ds^{2}} + \frac{B^{\prime }}{2A}\left( \frac{dt}{ds}\right) ^{\!2}\!\!+%
\frac{\dot{A}}{A}\frac{dt}{ds}\frac{dr}{ds}+\frac{A^{\prime }}{2A}\left( 
\frac{dr}{ds}\right) ^{2}\!\!\!=\!\frac{r}{A}\left( \frac{d\varphi }{ds}\right)^{2} ,  \label{eq3}
\end{equation}%
\begin{equation}
\frac{d^{2}\varphi }{ds^{2}}+\frac{2}{r}\frac{dr}{ds}\frac{d\varphi }{ds}=0,
\label{eq4}
\end{equation}%
where 
$(\dot{\phantom{.}})=\partial /\partial t$, $\left( ^{\prime }\right)
=\partial /\partial r$. From Eqs. (\ref{eq4}) and (\ref{eq1b}) it follows
that%
\begin{eqnarray}
\frac{d\varphi }{ds}&=&\frac{J}{r^{2}}, \label{eq5}\\  
B\left( \frac{dt}{ds}\right) ^{2}-&A&\left( \frac{dr}{ds}\right) ^{2}=1+\frac{J^{2}}{r^{2}},  \label{eq6}
\end{eqnarray}%
where $J=const$ is the particle's angular momentum.

Now suppose for a moment that the trajectory of the particle is known.
Passing from variable $s$ to variable $t$ and denoting%
\begin{equation}
Y(t)=B(t,r(t))\left( ds/dt\right) ^{-1},  \label{eq7}
\end{equation}%
from Eq. (\ref{eq6}) we get%
\begin{equation}
\left( \frac{dr}{dt}\right) ^{2}=\frac{B}{A}\left[ 1-\frac{B}{Y^{2}}\left( 1+%
\frac{J^{2}}{r^{2}}\right) \right] .  \label{eq8}
\end{equation}%
From Eq. (\ref{eq2}) with (\ref{eq8}), the equation for $Y^{2}$ follows:%
\begin{equation}
\frac{dY^{2}}{dt}=\left( \frac{\dot{B}}{B}-\frac{\dot{A}}{A}\right) Y^{2}+%
\frac{\dot{A}B}{A}\left( 1+\frac{J^{2}}{r^{2}}\right) .  \label{eq9}
\end{equation}%
Given that $d/ds=\left( Y/B\right) d/dt$ and using Eqs. (\ref{eq8}) and (\ref%
{eq9}), we find%
\begin{equation}
\frac{d^{2}r}{ds^{2}}=\frac{Y^{2}}{B^{2}}\left[ \frac{d^{2}r}{dt^{2}}-\frac{%
\dot{B}}{2B}\frac{dr}{dt}-\frac{B^{\prime }}{B}\left( \frac{dr}{dt}\right)
^{2}-\frac{\dot{A}}{2B}\left( \frac{dr}{dt}\right) ^{3}\right] .
\label{eq10a}
\end{equation}%
Substituting (\ref{eq10a}) into Eq. (\ref{eq3}) and taking into account 
(\ref{eq5}), (\ref{eq7}), and (\ref{eq8}), we finally obtain%
\begin{eqnarray}
\frac{d^{2}r}{dt^{2}}\!&+&\!\left( \frac{\dot{A}}{A}\!-\!\frac{\dot{B}}{2B}\right) 
\frac{dr}{dt}+\left( \frac{\gamma (r)}{r}+\frac{A^{\prime }}{2A}\!-\!\frac{%
B^{\prime }}{B}\right) \left( \frac{dr}{dt}\right) ^{2}\quad\quad \nonumber \\
&-&\frac{\dot{A}}{2B}\left( \frac{dr}{dt}\right) ^{3}\!=\!\frac{\gamma (r)}{r}%
\frac{B}{A}\!-\!\frac{B^{\prime }}{2A},  \label{eq11}
\end{eqnarray}
\begin{equation}
\left( \frac{d\varphi }{dt}\right) ^{2}=\frac{\gamma (r)}{r^{2}}\left[
B-A\left( \frac{dr}{dt}\right) ^{2}\right] ,  \label{eq12}
\end{equation}
where%
\begin{equation}
\gamma (r)=\frac{J^{2}/r^{2}}{1+J^{2}/r^{2}},\qquad 0\leqslant \gamma <1.
\label{eq13}
\end{equation}%
For Eqs. (\ref{eq11}) and (\ref{eq12}) we choose the initial conditions%
\begin{equation}
r(0)=r_{0},\quad \left( dr/dt\right) _{t=0}=0,\quad \varphi (0)=0.
\label{eq14}
\end{equation}%
The equations (\ref{eq11})-(\ref{eq14}) completely determine the time dependence 
of the coordinates of the moving particle. For radial motions we set 
$\gamma =0$. For orbital motions, the requirement for the absence of a
constant component on the right hand side of Eq. (\ref{eq11}) leads to the
condition%
\begin{equation}
\gamma (r_{0})=\frac{r_{0}}{2}\frac{\overline{B^{\prime }(t,r_{0})/A(t,r_{0})%
}}{\overline{B(t,r_{0})/A(t,r_{0})}},  \label{eq15}
\end{equation}%
where the bar means the average over the period of oscillations of the
gravitational field. This condition gives the relation between $r_{0}$ and
the angular momentum $J$ at which, in the case of a static mass
distribution, the particle moves along a circular orbit of radius $r_{0}$.
Of course, this is only possible if the right hand side of Eq. (\ref{eq15})
is positive and less than one. In this case, for a given $J$, the value of $%
r_{0}$ is found by solving Eq. (\ref{eq15}). Alternatively, one can choose $%
r_{0}$ and calculate $\gamma _{0}=\gamma (r_{0})$ using Eq. (\ref{eq15}).
This gives $J^{2}=r_{0}^{2}\gamma _{0}/\left( 1-\gamma _{0}\right) $ for the
squared momentum, and hence $\gamma (r)$ involved in Eqs. (\ref{eq11})-(\ref{eq13}) 
can be written as%
\begin{equation}
\gamma (r)=\left[ 1+\frac{1-\gamma _{0}}{\gamma _{0}}\left( \frac{r}{r_{0}}%
\right) ^{2}\right] ^{-1}.  \label{eq15a}
\end{equation}

\subsection{Weak field approximation}

Now let us assume that the oscillating lump of DM has a low density, so that
its gravitational field is weak everywhere on the particle trajectory. It
follows that%
\begin{equation}
A=1-2\psi +O(\varkappa ^{2}),\qquad B=1+2\chi +O(\varkappa ^{2}),
\label{eq16}
\end{equation}%
where $\psi (t,r)$ and $\chi (t,r)$ are small time-periodic functions of
order $\varkappa $, $\varkappa \ll 1$ being a dimensionless small parameter
proportional to the gravitational constant $G$. With finite motions of
particles in such a gravitational field, the particle velocities are small.
Taking this into account and using (\ref{eq16}), from Eq. (\ref{eq11}) we
obtain%
\begin{equation}
\frac{d^{2}r}{dt^{2}}-\left( 2\dot{\psi}(t,r)+\dot{\chi}(t,r)\right) \frac{dr%
}{dt}=\frac{\gamma (r)}{r}-\chi ^{\prime }(t,r).  \label{eq17}
\end{equation}%
When deriving this equation, we neglected the terms of the orders $\varkappa
^{2}/R_{g},\;\left( \varkappa ^{2}/T_{g}\right) dr/dt,\;\left( \varkappa
/R_{g}\right) \left( dr/dt\right) ^{2},$\ and $\left( \varkappa
/T_{g}\right) \left( dr/dt\right) ^{3}$, where $R_{g}$ is the characteristic
radius of the gravitating lump, $T_{g}$ is the oscillation period of the
gravitational field. In the same approximation, Eq. (\ref{eq12}) becomes%
\begin{equation}
\left( \frac{d\varphi }{dt}\right) ^{2}=\frac{\gamma (r)}{r^{2}}\left[
1+2\chi (t,r)\right] ,  \label{eq18}
\end{equation}%
and condition (\ref{eq15}) takes the form%
\begin{equation}
\gamma (r_{0})=r_{0}\bar{\chi}^{\prime }(r_{0}),  \label{eq19}
\end{equation}%
where we denote $\bar{\chi}(r_{0})=\overline{\chi (t,r_{0})}$.

Note that Eq. (\ref{eq17}) is generally nonlinear. Nevertheless, there are
two types of finite motions for which one can restrict oneself to a linear
analysis. Consider, at first, small radial oscillations near the center.
Setting $\gamma =0$, we expand $\psi (t,r)$ and $\chi (t,r)$ at $r=0$, taking
into account that $\psi ^{\prime }(t,0)=\chi ^{\prime }(t,0)=0$ due to the
assumed smoothness of the gravitational field. Neglecting the terms of the
orders $\left( \varkappa /T_{g}\right) \left( r/R_{g}\right) ^{2}dr/dt$ and $%
\varkappa r^{2}/R_{g}^{3}$, we obtain%
\begin{equation}
\frac{d^{2}r}{dt^{2}}-\left( 2\dot{\psi}(t,0)+\dot{\chi}(t,0)\right) \frac{dr%
}{dt}+\chi ^{\prime \prime }(t,0)r=0.  \label{eq20}
\end{equation}%
Substitution%
\begin{equation}
r(t)=u(t)\exp \left( \psi (t,0)+\frac{1}{2}\chi (t,0)\right)  \label{eq21}
\end{equation}%
transforms Eq. (\ref{eq20}) into the equation%
\begin{equation}
\frac{d^{2}u}{dt^{2}}+\left( \ddot{\psi}(t,0)+\frac{1}{2}\ddot{\chi}%
(t,0)+\chi ^{\prime \prime }(t,0)\right) u=0,  \label{eq22}
\end{equation}%
where in the brackets the terms of the order $\varkappa ^{2}/T_{g}^{2}$ were
neglected. The remaining terms are $T_{g}$-periodic. Equations of this type are usually
called Hill equations. They have numerous applications in physics and
technology (see, e.g., \cite{Stoker}). The properties of solutions to the
Hill equations depend significantly on the parameters involved in the
equation. In the parameter space, there are regions in which solutions grow
exponentially with time (resonant zones) and regions in which solutions are
bounded (nonresonant zones). The Floquet theory of the Hill equation is
presented, e.g., in the book \cite{Magnus}.

Now consider the motion along a trajectory close to circular. In this case
we set%
\begin{equation}
r(t)=r_{0}\left( 1+\eta (t)\right) ,  \label{eq23}
\end{equation}%
where $\eta (t)$ is small. Making expansions in Eq. (\ref{eq17}) at $%
r=r_{0} $ and taking into account (\ref{eq19}), we obtain%
\begin{eqnarray}
\frac{d^{2}\eta }{dt^{2}}&-&\left( 2\dot{\psi}(t,r_{0})+\dot{\chi}%
(t,r_{0})\right) \frac{d\eta }{dt}\nonumber\\
&+&\left( \chi ^{\prime \prime }(t,r_{0})+\frac{3\gamma (r_{0})}{r_{0}^{2}}\right) \eta
=-\frac{1}{r_{0}}\tilde{\chi}^{\prime }(t,r_{0}),  \label{eq24}
\end{eqnarray}%
where $\tilde{\chi}(t,r_{0})=\chi (t,r_{0})-\bar{\chi}(r_{0})$ and the terms
of the orders $\left( r_{0}/R_{g}\right) \left( \varkappa \eta /T_{g}\right)
d\eta /dt$,$\;\varkappa ^{2}\eta /R_{g}^{2}$,\ $\varkappa \eta ^{2}/\left(
r_{0}R_{g}\right) $, and $r_{0}\varkappa \eta ^{2}/R_{g}^{3}$ were
neglected. Finally, substitution%
\begin{equation}
\eta (t)=u(t)\exp \left( \psi (t,r_{0})+\frac{1}{2}\chi (t,r_{0})\right)
\label{eq25}
\end{equation}%
results in the nonhomogeneous Hill equation%
\begin{eqnarray}
&&\frac{d^{2}u}{dt^{2}}+\left( \ddot{\psi}(t,r_{0})+\frac{1}{2}\ddot{\chi}%
(t,r_{0})+\chi ^{\prime \prime }(t,r_{0})+\frac{3\gamma (r_{0})}{r_{0}^{2}}%
\right) u  \nonumber \\
&=&-\frac{1}{r_{0}}\tilde{\chi}^{\prime }(t,r_{0}),  \label{eq26}
\end{eqnarray}%
where we neglected the terms of the orders $\varkappa ^{2}/T_{g}^{2}$ and 
$\varkappa ^{2}/\left(r_{0}R_{g}\right) $ in the brackets and on the right hand side, respectively.

The nonhomogeneous Hill equation has also been studied in the literature
(see, e.g., \cite{Rodriguez} and references therein), but not in as much
detail as its homogeneous counterpart. For our analysis, it is only
important that the periodic forcing term on the right hand side of Eq. (\ref{eq26}) 
does not affect the location of the boundaries of the resonant zones \cite{Kotowski, Slane}.

\section{Motions of the test particle in a time-periodic spherically
symmetric scalar field}

As examples, we consider radial and orbital motions of the test particle in
the self-gravitating real scalar field with the potential%
\begin{equation}
U(\phi )=\frac{m^{2}}{2}\phi ^{2}\left( 1-\ln \frac{\phi ^{2}}{\sigma ^{2}}%
\right) ,  \label{eq27}
\end{equation}%
where $\sigma $ is the characteristic magnitude of the field, $m$ is the
mass (in units $\hbar =c=1)$. This potential is singular: its second
derivative tends to infinity as $\phi $ passes through zero. 
Originally, such potentials were considered in quantum field theory in
connection with dilatation covariance of relativistic field equations 
\cite{Rosen}. Later in Refs. \cite{Birula, Birula2} it was shown that the
requirements of separability of non-interacting quantum subsystems and the
validity of Planck's relation $E=\hbar \omega$ for stationary states
uniquely determine the logarithmic potential (\ref{eq27}) (with a complex
wavefunction) in the Schr\"{o}dinger-Pauli and Klein-Gordon-type equations.
Also, when taking into account quantum corrections, such potentials
naturally appear in the inflationary cosmology \cite{Linde, Barrow, Enqvist2}%
, as well as in some supersymmetric extensions of the Standard Model (flat
direction potentials in the gravity mediated supersymmetric breaking
scenario) \cite{Enqvist, Kasuya}. It is remarkable that in the Minkowski
spacetime the potential (\ref{eq27}) admits a whole family of exact
solutions of the Klein-Gordon equation in the form $\phi =a(t)w(\mathbf{r})$, 
describing multidimensional localized time-periodic configurations of a
real scalar field, the pulsons (oscillons) \cite{Marques, Bogolubsky}.
Moreover, it turned out that potential (\ref{eq27}) is the only one that
allows such solutions to exist \cite{Maslov}. It was also shown that pulsons
can arise due to the fragmentation of a homogeneous scalar background
oscillating around the local minimum of the potential (\ref{eq27}) 
\cite{Koutvitsky0}. Stability analysis of the real pulsons have shown that,
despite the absence of a global charge, there are values of oscillation
amplitudes at which they are long-lived objects that retain their
periodicity for a long time \cite{Koutvitsky5, Koutvitsky0, Ibe}.

The above mentioned unique properties of the logarithmic potential motivate
us to use this field model to test our approach. To do this, we need to take
into account the effects of self-gravity. The corresponding solution of the
Einstein--Klein--Gordon system was found in Ref. \cite{Koutvitsky2} by the
Krylov-Bogoliubov method.
This solution describes a self-gravitating field
lump of an almost Gaussian shape that pulsates in time. In the weak field
approximation, the corresponding metric functions $A(t,r)$ and $B(t,r)$ can
be written as (\ref{eq16}), where
\begin{equation}\label{eq28}
\psi (t,r)\!=\frac{\varkappa }{2}\!\left[a^{2}\rho ^{2}\!+\!V_{\max }\!\left( 1\!-\!\frac{\sqrt{\pi }\,%
\mathrm{erf}\,\rho }{2\rho }e^{\rho ^{2}}\!\right) \!\right]\,e^{3-\rho ^{2}},  \\
\end{equation}
\vspace{-12pt}
\begin{equation} \label{eq29}
\chi (t,r)\!=\!-\frac{\varkappa }{2}\!\left[ a^{2}\ln a^{2}\!+\!V_{\max }\!\left( 1\!+\!\frac{\sqrt{\pi }\,%
\mathrm{erf}\,\rho }{2\rho }e^{\rho ^{2}}\!\right) \!\right]\,e^{3-\rho ^{2}}, 
\end{equation}
$\tau =mt$, $\rho =mr$, $\varkappa =4\pi G\sigma ^{2}\ll 1$ ($G$ is the
gravitational constant). The function $a(\theta (\tau ))$ oscillates in the
range $-a_{\max }\leqslant a(\theta )\leqslant a_{\max }$\ $\left( 0<a_{\max
}<1\right) $ in the local minimum of the potential $V(a)$,%
\begin{equation}
a_{\theta \theta }=-dV/da,\qquad V(a)=(a^{2}/2)\left( 1-\ln a^{2}\right) ,
\label{eq30}
\end{equation}%
where $V_{\max }=V(a_{\max })$, $d\theta /d\tau =1+\varkappa \Omega
+O(\varkappa ^{2})$, and the constant $\varkappa \Omega $ is the frequency
correction due to gravitational effects (see Ref. \cite{Koutvitsky2} for
details). The period (in $\theta $) of these oscillations can be
approximated by%
\begin{equation}
T\approx 2\pi \left( 1-\ln a_{\max }^{2}\right) ^{-1/2}\qquad \left( a_{\max
}^{2}\ll 1\right) ,  \label{eq31}
\end{equation}%
\begin{equation}
T\approx 2\sqrt{2}\ln \left( 1-a_{\max }^{2}\right) ^{-1}\qquad \left(
1-a_{\max }^{2}\ll 1\right) .  \label{eq32}
\end{equation}

The energy density of the field lump we are considering is concentrated on
the characteristic scale $r\sim R_{g}=m^{-1}$ and decays as 
\begin{equation}
T{{_{0}^{0}}}\sim m{{^{2}}}\sigma ^{2}a^{2}(\theta )\rho ^{2}e^{3-\rho ^{2}}{%
\qquad }\left( \rho =r/R_{g}\gg 1\right) .  \label{eq33}
\end{equation}%
As seen from Eqs. (\ref{eq28}) and (\ref{eq29}), at large distances from the
lump the gravitational field turns into the static Schwarzschild field, in
accordance with the Birkhoff theorem (see, e.g., \cite{Weinberg}). However,
inside the lump the gravitational field oscillates with the period $%
T_{g}=[2m(1+\varkappa \Omega )]^{-1}T$ (with respect to $t$). Assuming no
direct interaction with the scalar field, let us consider the radial and
orbital motions of a test particle in this gravitational field.

\subsection{Radial motion}

As has been shown, small radial oscillations of a test particle are
generally described by Eqs. (\ref{eq21}) and (\ref{eq22}). Considering Eq. (%
\ref{eq22}), we can put $d/dt\approx m\,d/d\theta $ with the accepted
accuracy, and $d/dr=m\,d/d\rho $. Using Eqs. (\ref{eq28})-(\ref{eq30}) we
obtain%
\begin{equation}
\rho =v\,\exp \left( -\frac{\varkappa }{4}e^{3}\left( a^{2}\ln
a^{2}+2V_{\max }\right) \right) ,  \label{eq34}
\end{equation}%
\begin{eqnarray}
\frac{d^{2}v}{d\theta ^{2}}&+&\varkappa e^{3} \Bigg[ \frac{3}{2}a^{2} - a^{2}\ln a^{2}\left(\frac{1}{2}+\ln a^{2}\right) \nonumber\\
&-& V_{\max }\left( \frac{5}{3}+\ln a^{2}\right) \Bigg]v=0.  \label{eq35}
\end{eqnarray}
Eq. (\ref{eq35}) is the singular Hill equation because the expression in the
square brackets is periodic in $\theta $ and tends to infinity as $a(\theta )
$ passes through zero. For this equation, Fig. 1 shows the parametric
resonance zones on the $\varkappa -a_{\max }^{2}$ plane. 

\begin{figure}[t]
\includegraphics[width=8.5cm]{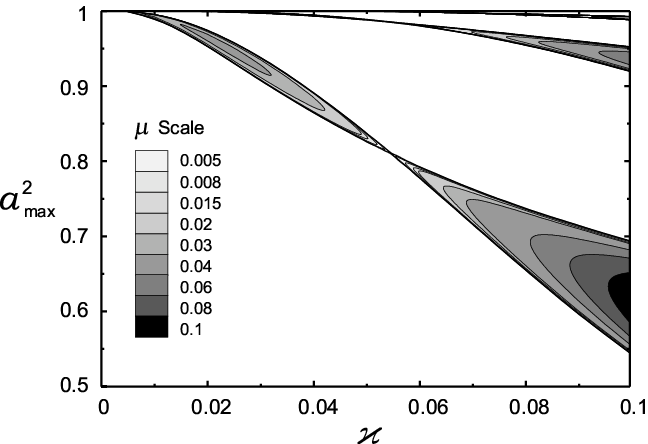}
\caption{Resonance zones and Floquet exponent $\mu$ for Eq. (\ref{eq35})}
\end{figure}

We solved Eqs. (\ref{eq30}), (\ref{eq34}), and (\ref{eq35}) numerically in both 
resonant and nonresonant zones using initial conditions%
\begin{equation}
a(0)=a_{\max },\;\;\;  a_{\theta }(0)=0,\;\;\;  \rho (0)=\rho _{0},\;\;\;  \rho
_{\theta }(0)=0,  \label{eq36}
\end{equation}%
\begin{equation}
v(0)=\rho _{0}\exp \left( \frac{\varkappa }{4}e^{3}a_{\max }^{2}\right)
,\;\;\; v_{\theta }(0)=0.  \label{eq37}
\end{equation}%
For comparison, using the same initial conditions (\ref{eq36}), we
numerically solved the exact equation (\ref{eq11}) with $\gamma =0$ and
metric functions $A(t,r)$ and $B(t,r)$ given by formulas (\ref{eq16}), 
(\ref{eq28}), and (\ref{eq29}), The results are shown in Fig. 2.

\begin{widetext}

\begin{center}
\begin{figure}[h]
\includegraphics[width=14cm]{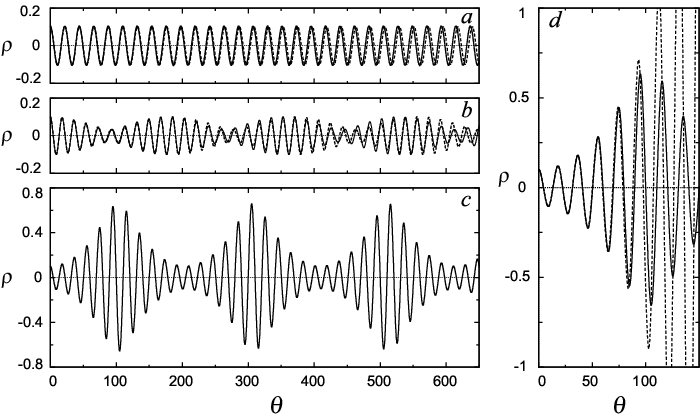}
\caption{Solutions of the equations (\ref{eq30}), (\ref{eq34}) and (\ref{eq35}) far from the resonance zone ($a$), 
in the vicinity of the resonance zone ($b$), and at the center of the resonance zone ($c$), 
$\rho_0 = 0.1$, $\varkappa=0.01$, $a_{\rm max}=0.9\; (a),\: 0.995\;(b),\: 0.997\; (c)$. 
The results are shown in dashed lines.
The corresponding solutions of the nonlinear Eq. (\ref{eq11}) for the same parameters are shown in solid lines. Panel ($d$) depicts 
the comparison of the initial fragment of ($c$) with the exponentially growing solution obtained in the linear approximation} 
\end{figure}
\end{center}
\vspace{-24pt}

\end{widetext}

\subsection{Orbital motion}
In this case, the particle trajectory, close to circular, is generally
described by Eqs. (\ref{eq23}), (\ref{eq25}), (\ref{eq26}), (\ref{eq18}),
and (\ref{eq19}). Using Eqs. (\ref{eq28})-(\ref{eq30}), with the accepted
accuracy, we obtain%
\begin{eqnarray}
&&\eta =u\,\exp S\left( a,\rho _{0}\right),  \label{eq38}\\
&&\frac{d^{2}u}{d\theta ^{2}}+\Big\{\, \frac{\gamma _{0}}{\rho _{0}^{2}}+\frac{\varkappa }{2}\Big[\, 2V_{\max }\left( 1-\ln a^{2}\right)+ \left( 3-2\rho _{0}^{2}\right)a^{2} \nonumber \\
&& -a^{2}\ln a^{2}\left(1+2\ln a^{2}\right)+ 4\,\,\overline{a^{2}\ln a^{2}}\,\Big]e^{3-\rho _{0}^{2}} \Big\} u \nonumber \\
&& = -\varkappa \,\widetilde{a^{2}\ln a^{2}}\,e^{3-\rho _{0}^{2}},{}  \label{eq39}
\end{eqnarray}
where%
\begin{eqnarray}
&&\!\!\!\!S\left( a,\rho \right) =\frac{\varkappa }{4}e^{3-\rho ^{2}}\times {}\nonumber\\
&&\!\!\!\!\left[ 2a^{2}\rho^{2}\!-a^{2}\ln a^{2}\!+V_{\max }\!\left( 1-\frac{3\sqrt{\pi }\,\mathrm{erf}\,\rho 
}{2\rho }e^{\rho ^{2}}\right) \right]\!,  \label{eq40}
\end{eqnarray}%
\begin{eqnarray}
&&\!\!\!\!\!\!\gamma _{0}=\varkappa \rho _{0}^{2}e^{3-\rho _{0}^{2}}\times {}\nonumber\\
&&\!\!\!\!\!\!\left\{ \overline{a^{2}\ln a^{2}}\!+\!V_{\max }\!\left[ 1\!-\!\frac{1}{2\rho _{0}^{2}}\left( 1\!-\!\frac{%
\sqrt{\pi }\mathrm{erf}\,\rho _{0}}{2\rho _{0}}e^{\rho _{0}^{2}}\right) %
\right] \right\}\!,  \label{eq41}
\end{eqnarray}%
and $\widetilde{a^{2}\ln a^{2}}=a^{2}\ln a^{2}-\overline{a^{2}\ln a^{2}}$.
Eq. (\ref{eq39}) is nonhomogeneous singular Hill equation. The resonance 
zones of this equation are depicted in Fig. 3 on the $\varkappa -a_{\max}^{2}$ 
plane with fixed $\rho _{0}$. 
\begin{figure}[t]
\includegraphics[width=8.5cm]{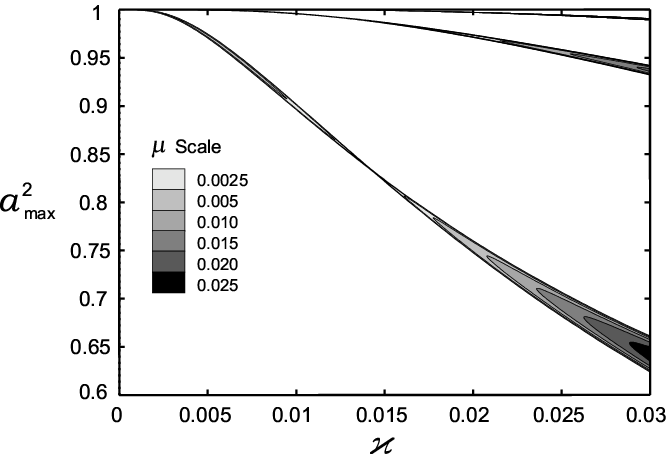}
\caption{Resonance zones and Floquet exponent $\mu$ for Eq. (\ref{eq39}) with $\rho _{0}=0.3$}
\end{figure}
We solved Eqs. (\ref{eq30}), (\ref{eq38}), and (\ref{eq39}) numerically using the initial conditions%
\begin{equation}
a(0)=a_{\max },\quad a_{\theta }(0)=0,\quad \eta (0)=0,\quad \eta _{\theta}(0)=0,  \label{eq42}
\end{equation}%
\begin{equation}
u(0)=0,\quad u_{\theta }(0)=0.  \label{eq43}
\end{equation}%
Further, assuming that $\gamma (r)$, $A(t,r)$, and $B(t,r)$ are given by
formulas (\ref{eq15a}), (\ref{eq16}), (\ref{eq28}), and (\ref{eq29}), we
numerically solved Eq. (\ref{eq11}) to find $\eta =\left( \rho -\rho
_{0}\right) /\rho _{0}$ with the same initial conditions (\ref{eq42}). The
results can be compared in Fig. 4. In addition, we show the evolution of the
orbital trajectory obtained by integrating Eq. (\ref{eq12}).

\section{Discussion}

Thus, based on the geodesic equation, we have derived the nonlinear system
determining the time dependence of the test particle coordinates in finite
motions in the oscillating spherically symmetric spacetime. In the weak
field approximation we have reduced the nonlinear equation (\ref{eq11}) for
the radial coordinate to linear equations, namely, to the Hill equation (\ref%
{eq22}) for the radial oscillations and the nonhomogeneous Hill equation (\ref%
{eq26}) for the oscillations near circular orbit, respectively. Using these
equations we studied resonance effects in radial and orbital motions of test
particles inside oscillating spherically symmetric field lump in the scalar
model with the logarithmic self-interaction (\ref{eq27}). Equations (\ref%
{eq22}) and (\ref{eq26}) then take the form (\ref{eq35}) and (\ref{eq39}),
respectively.

The resonant solutions of the Hill equation at large times $\theta $ have
the asymptotic form $\sim F(\theta )e^{\mu \theta }$, where $F(\theta )$ is
a $T$-periodic ($T/2$-periodic or $T/2$-antiperiodic) function, and the
Floquet exponent $\mu $ is positive in parametric resonance zones (see,
e.g., \cite{Magnus}). In the case of radial motions, the resonance zones of
Eq. (\ref{eq35}) are depicted in Fig. 1 on the $\varkappa -a_{\max }^{2}$
plane. We solved numerically Eq. (\ref{eq35}) at different points of this
plane and compared the solutions with the corresponding solutions of Eq. (%
\ref{eq11}) in the domain of sufficiently small $\varkappa $ and $\rho _{0}$
for which our approximation is valid. We have found that, far
from the resonance zones, the solutions are in good agreement with each
other over a sufficiently large time interval (see Fig. 2a, as an example).
It can be seen that the oscillations are practically sinusoidal. As you
approach (with fixed $\varkappa $) the resonance zone, these oscillations
become modulated, turning into beats (Fig. 2b). In the resonance zone, the
beats acquire a different character (Fig. 2c). As can be seen from Fig. 2d,
the increase in the oscillation amplitude in each beat is well approximated
by an exponent with a growth rate close to the Floquet exponent in the
corresponding solution of the Hill equation. This suggests a resonant
mechanism for the growth of oscillations. When the oscillation amplitude
reaches sufficiently large values, the high-order nonlinear terms in Eq. (%
\ref{eq11}) come into play, periodically suppressing the resonance and
limiting the amplitude of the beats.

Now consider the orbital motions. The resonance zones of Eq. (\ref{eq39})
are shown in Fig. 3 on the $\varkappa -a_{\max }^{2}$ plane for a fixed $%
\rho _{0}$. Far from resonance zones, the solutions of Eqs. (\ref{eq38})
and (\ref{eq39}) are in excellent agreement with solutions of Eq. (\ref{eq11}) 
(see Fig. 4a, where the curves in the inset practically coincide).
Near the resonant zones, the solutions gradually diverge with time (Fig.
4b). Nevertheless, it is clear that the solutions of Eqs. (\ref{eq38}) and (%
\ref{eq39}) here gives an acceptable description of the beats that have
appeared. However, in the resonance zones, the solutions diverge
significantly (see Fig. 4c,d). First, we have found that the resonance zones
of Eq. (\ref{eq11}) are located within the resonance zones of Eq. (\ref%
{eq39}), but they are much narrower. Secondly, the growth rate of resonant
solutions of Eq. (\ref{eq11}) is much less (more than three times at
maximum) than the Floquet exponent for the corresponding solutions of Eq. (%
\ref{eq39}). We believe that, as in the case of radial motions, the
high-order nonlinear terms in Eq. (\ref{eq11}) tame the resonance, which
results in limiting the growth of oscillations and appearance 
of the long-period beats (Fig. 4c). As expected, the
maximum values of the amplitude and period of the beats were observed at the
center of the resonance zone of Eq. (\ref{eq11}). In this regard, our
results are similar to those obtained in Ref. \cite{Boskovic} for the case
of orbital motion of a test particle inside an oscillating star with uniform
density. The role of nonlinear terms was also discussed in Ref. 
\cite{Cairncross} in the study of parametric resonance in Bose-Einstein
condensates.

\begin{widetext}

\begin{center}
\begin{figure}[ht]\label{Figure4}
\includegraphics[width=14cm]{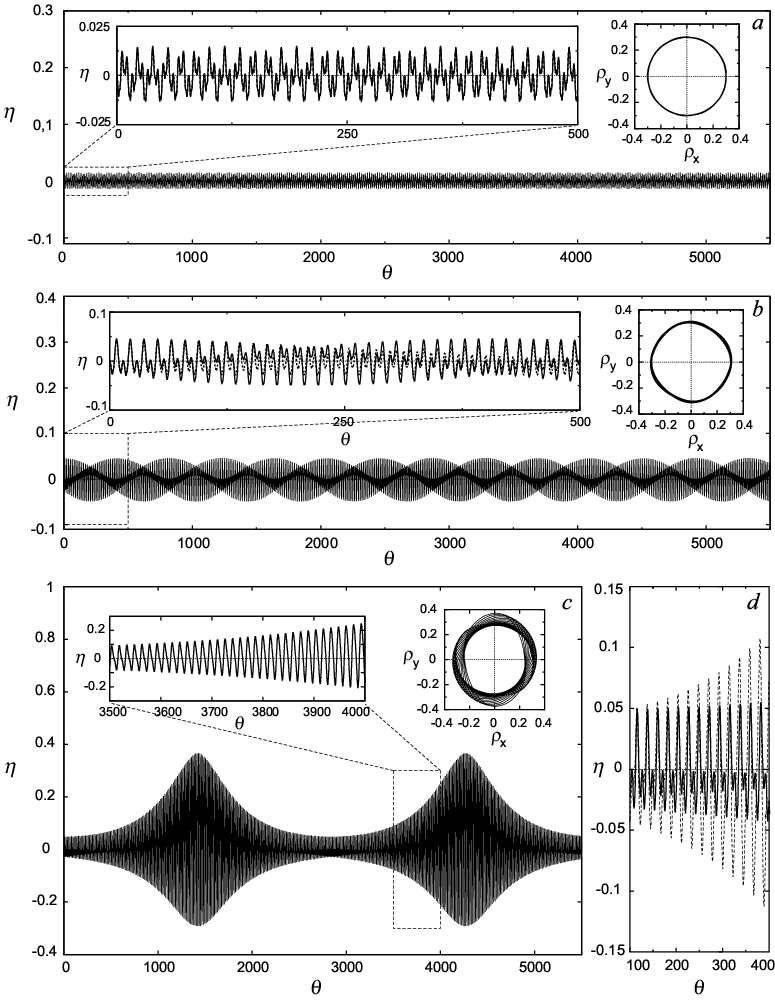}
\caption{Solutions of the equations (\ref{eq30}), (\ref{eq38}), 
and (\ref{eq39}) far from the resonance zone ($a$), in the vicinity of the resonance zone ($b$), 
and at the center of the same zone ($c$). The results of integration are shown in dashed lines,
$\rho_0 = 0.3$, $\varkappa=0.005$, $a_{\rm max}=0.950\; (a),\: 0.985\;(b),\: 0.98625\; (c)$. The %
corresponding solutions of the nonlinear system (\ref{eq11}), (\ref{eq12}) are shown in solid lines.
Panel ($d$) presents the comparison of the initial fragment of ($c$) with the rapidly growing solution 
obtained in the linear approximation  (see Eqs. (\ref{eq38}) and (\ref{eq39})). 
The nearly circular particle orbits shown in the right insets of panels $(a), (b)$ and $(c)$ 
correspond to the oscillation curves depicted in the left insets } 
\end{figure}
\end{center}

\end{widetext}

Summing up, we can conclude that in the absence of resonance, the finite
motions of a test particle are well described in the linear approximation by
the Hill equations. In the resonance zones, the nonlinearities suppress the
resonance and limit the growth of the oscillation amplitude. Here, the
resonant effects manifest themselves only in a change in the shape of the
beats and must be described within the framework of the original nonlinear
system. 

When applied to the motion of short-period stars near the center of the Galaxy, 
where the density of dark matter is presumably greatest, the resonance phenomena 
we are discussing can manifest themselves in the form of characteristic oscillations 
(with the specific beats) of the star's radial velocity and as slow variations 
of the orbital eccentricity with a beating period. Because these effects are small, 
very precise and long-term measurements of the orbital parameters and radial velocity 
are required to estimate the deviation of the star's radial velocity from 
that expected for stable orbital motion.
 
As concerns the central star cluster, however, so far this deviation has been 
evaluated  only for the most well-studied short-period star S0-2 in the search 
for its binarity \cite{Ghez}. In this research, the radial velocity variations from 
the S0-2 orbital model have been studied using data from several dozen of radial 
velocity measurements but no significant periodic signals have been detected 
at the current level of measurement accuracy. The expected uncertainties 
with Extremely Large Telescope will be lowered by one order of magnitude or so, 
but as it was estimated in \cite{Boskovic}, only reduction of this uncertainties 
by two orders of magnitude can reveal fluctuations caused by the above 
mentioned effects of axion DM instability.

We hope that upcoming long-term spectroscopic studies of stars in the center 
of the Galaxy using new generation telescopes will provide more accurate data 
and confirm the ultra-light axion nature of dark matter or impose new constraints 
on the models under consideration.

\section*{Acknowledgment}
We would like to thank the referee for useful comments which contributed to the improvement of the paper.

\end{document}